\documentclass[12pt]{article}
\usepackage{fullpage}
\usepackage{epsf}
\usepackage{epsfig}
\usepackage{amsthm}
\usepackage{amsfonts}
\usepackage{color}
\usepackage{amsmath}
\usepackage{natbib}

\begin{document}

\newcommand{\bm}[1]{\mbox{\boldmath $#1$}}
\newcommand{\mb}[1]{\mathbf{#1}}
\newcommand{\bE}[0]{\mathbb{E}}
\newcommand{\bP}[0]{\mathbb{P}}
\newcommand{\ve}[0]{\varepsilon}
\newcommand{\mN}[0]{\mathcal{N}}
\newcommand{\iidsim}[0]{\stackrel{\mathrm{iid}}{\sim}}
\newcommand{\NA}[0]{{\tt NA}}

\title{\vspace{-0.5cm} Particle learning of Gaussian process 
models for sequential design and optimization}
\author{Robert B.~Gramacy\\
  Statistical Laboratory\\
  University of Cambridge\\
  {\tt bobby@statslab.cam.ac.uk} \and
  Nicholas G.~Polson\\
  Booth School of Business\\
  University of Chicago\\
  {\tt ngp@chicagobooth.edu}}
\date{}

\maketitle

\begin{abstract}
  We develop a simulation-based method for the online updating of
  Gaussian process regression and classification models.  Our method
  exploits sequential Monte Carlo to produce a fast sequential design
  algorithm for these models relative to the established MCMC
  alternative.  The latter is less ideal for sequential design since
  it must be restarted and iterated to convergence with the inclusion
  of each new design point.  We illustrate some attractive ensemble
  aspects of our SMC approach, and show how active learning heuristics
  may be implemented via particles to optimize a noisy function or to
  explore classification boundaries online.

  \bigskip
  \noindent {\bf Key words:} Sequential Monte Carlo,
  Gaussian process, nonparametric regression and classification,
  optimization, expected improvement, sequential design, entropy
\end{abstract}

\section{Introduction}
\label{sec:intro}

The Gaussian process (GP) is by now well established as the backbone
of many highly flexible and effective nonlinear regression and
classification models \citep[e.g.,][]{neal:1998,rasmu:will:2006}.  One
important application for GPs is in the sequential design of computer
experiments \citep{sant:will:notz:2003} where designs are built up
iteratively: choose a new design point $x$ according to some criterion
derived from a GP surrogate model fit; update the fit conditional on
the new pair $(x, y(x))$; and repeat.  The goal is to keep designs
small in order to save on expensive simulations of $y(x)$.  By ``fit''
we colloquially mean: samples obtained from the GP posterior via MCMC.
While it is possible to choose each new design point via full
utility-based design criterion \citep[e.g.,][]{muller:sanso:dei:2004},
this can be computationally daunting even for modestly sized designs.
More thrifty {\em active learning} (AL) criterion such as ALM
\citep{mackay:1992} and ALC \citep{cohn:1996} can be an effective
alternative.  These were first used with GPs by \cite{seo00}, and have
since been paired with a non-stationary GP to design a rocket booster
\citep{gra:lee:2009}.

Similar AL criteria are available for other sequential design tasks.
Optimization by {\em expected improvement}
\citep[EI,][]{jones:schonlau:welch:1998} is one example.
\cite{tadd:lee:gray:grif:2009} used an embellished EI with a
non-stationary GP model and MCMC inference to determine the optimal
robust configuration of a circuit device.  In the classification
setting, characteristics like the predictive entropy
\citep{joshi:2009} can be used to explore the boundaries between
regions of differing class label in order to maximize the information
obtained from each new $x$.  The thrifty nature of AL and the
flexibility of the GP is a favorable marriage, indeed.  However, a
drawback of batch MCMC-based inference is that it is not tailored to
the online nature of sequential design.  Except to guide the
initialization of a new Markov chain, it is not clear how fits from
earlier iterations may re-used in search of the next $x$. So after the
design is augmented with $(x,y(x))$ the MCMC must be restarted and
iterated to convergence.

In this paper we propose to use a sequential Monte Carlo (SMC)
technique called {\em particle learning} (PL) to exploit the
analytically tractable (and Rao--Blackwellizable) GP posterior {\em
  predictive} distribution in order to obtain a quick update of the GP
fit after each sequential design iteration.  We then show how some key
AL heuristics may be efficiently calculated from the particle
approximation.  Taken separately, SMC/PL, GPs, and AL, are by now well
established techniques in their own right.  Our contribution lies in
illustrating how together they can be a potent mixture for sequential
design and optimization under uncertainty.

The remainder of the paper is outlined as follows. Section
\ref{sec:gp} describes the basic elements of GP modeling.  Section
\ref{sec:smc} reviews SMC and PL, highlighting the strengths of PL in
our setting.  Section \ref{sec:plgp} develops a PL implementation for
GP regression and classification, with illustrations and comparisons
to MCMC.  We show how fast updates of particle approximations may be
used for AL in optimization and classification in Section
\ref{sec:sd}, and we conclude with a discussion in Section
\ref{sec:discuss}.  Software implementing our methods, and the
specific code for our illustrative examples, is available in the {\tt
  plgp} package \citep{plgp} for {\sf R} on CRAN.

\subsection{Gaussian process priors for regression and classification}
\label{sec:gp}

A GP prior for functions $Y\!:\mathbb{R}^p \rightarrow \mathbb{R}$,
where any finite collection of outputs are jointly Gaussian
\citep{stein:1999}, is defined by its mean $\mu(x) = \bE\{Y(x)\} =
f(x)^\top \beta$ and covariance $C(x,x') = \bE\{[Y(x) - \mu(x)][(Y(x')
- \mu(x')]^\top]\}$.  Often the mean is linear in the inputs ($f(x) =
[1,x]$) and $\beta$ is an unknown $(p+1) \times 1$ parameter vector.
Typically, one separates out the variance $\sigma^2$ in $C(x,x')$ and
works with correlations $K(x,x') = \sigma^{-2} C(x,x')$ based on
Euclidean distance and a small number of unknown parameters; see,
e.g., \citet{abraham:1997}.  In the classical inferential setting we
may view the GP ``prior'' as a choice of model function, thus
resembling a likelihood.  

In the regression problem, the likelihood of data $D_N = (X_N, Y_N)$,
where $X_N$ is a $N\times p$ design matrix and $Y_N$ is a $N \times 1$
response vector, is multivariate normal (MVN) for $Y_N$ with mean
$\mu(X_N) = F_N \beta$, where $F_N$ is a $N \times (p+1)$ matrix that
contains $f(x_i)^\top$ in its rows, and covariance $\Sigma(X_N) =
\sigma^2 K_N$, where $K_N$ is the $N \times N$ covariance matrix with
$(ij)^{\mathrm{th}}$ entry $K(x_i, x_j)$. To reduce clutter, we shall
drop the $N$ subscript when the context is clear.  Conditional on $K$,
the MLE for $\beta$ and $\sigma^2$ is available in closed form via
weighted least squares.  The profile likelihood may be used to infer
the parameters to $K(\cdot, \cdot)$ numerically.

Bayesian inference may proceed by specifying priors over $\beta$,
$\sigma^2$, and the parameters to $K(\cdot, \cdot)$.  With priors
$\beta \propto 1$ and $\sigma^2 \sim \mathrm{IG}(a/2, b/2)$, the
marginal posterior distribution for $K(\cdot, \cdot)$, integrating
over $\beta$ and $\sigma^2$, is available in closed form
\citep[][Section A.2]{gramacy:2005}:
\begin{equation}
p(K|D) = p(K) \times \left(\frac{|V_\beta|}{|K|} \right)^{1/2} 
\times \frac{(b/2)^{\frac{a}{2}} 
\Gamma[(a+N-p)/2]}{(2\pi)^{\frac{N-p}{2}}\Gamma[a/2]} \times
\left(\frac{b + \psi}{2}\right)^{-\frac{a+N-p}{2}},
\label{eq:gpk}
\end{equation}
\begin{align}
\psi &= Y^\top K^{-1} Y - \tilde{\beta}^\top V_\beta^{-1} \tilde{\beta}, &
\tilde{\beta} &= V_\beta (F^\top K^{-1} Y), 
& V_\beta &= (F^\top K^{-1} F)^{-1}.
\label{eq:bmean}
\end{align}
It is possible to use a vague scale-invariant prior ($a,b=0$) for
$\sigma^2$.  In this case, the marginal posterior (\ref{eq:gpk}) is
proper as long as $N > p+1$.  Mixing is generally good for
Metropolis--Hastings (MH) sampling as long as $K(\cdot, \cdot)$ is
parsimoniously parameterized, $N$ is large, and there is a high
signal--to--noise ratio between $X$ and $Y$.  Otherwise, the posterior
can be multimodal \citep[e.g.,][]{warnes:ripley:1987} and hard to
sample.

Crucially for our SMC inference via PL [Section \ref{sec:plgp}], and
for our AL heuristics [Section \ref{sec:sd}], the fully marginalized
predictive equations for GP regression are available in closed form.
Specifically, the distribution of the response $Y(x)$ conditioned on
data $D$ and covariance $K(\cdot,\cdot)$, i.e., $p(y(x) | D, K)$, is
Student-$t$ with degrees of freedom $\hat{v} = N-p-1$,
\begin{align} 
  \mbox{mean} && \hat{y}(x|D, K) &= f(x)^\top \tilde{\beta} + k^\top(x)
  K^{-1}(Y-F\tilde{\beta}),
\label{eq:predgp} \\ 
\mbox{and scale} && 
 \hat{\sigma}^2(x|D, K) &=   
\frac{(b + \psi) [K(x, x) - k^\top(x)K^{-1} k(x)]}{a + \hat{\nu}}.
\label{eq:preds2}
\end{align}
where $k^\top(x)$ is the $N$-vector whose $i^{\mbox{\tiny th}}$
component is $K(x,x_i)$.

In the classification problem, with data $D = (X, C)$, where $C$ is a
$N \times 1$ vector of class labels $c_i \in \{1,\dots, M\}$, the GP
is used $M$-fold as a prior over a collection of $M \times N$ latent
variables $\mathcal{Y} = \{Y_{(m)}\}_{m=1}^M$, one set for each class.
For a particular class $m$, the generative model (or prior) over the
latent variables is MVN with mean $\mu_{(m)}(X)$ and variance
$\Sigma_{(m)}(X)$, as in the regression setup.  The class labels then
determine the likelihood through the latent variables under an
independence assumption so that $p(C_N | \mathcal{Y}) = \prod_{i=1}^N
p_i$, where $p_i = p(C(x_i) = c_i | \mathcal{Y}_i)$. \cite{neal:1998}
recommends a {\em softmax} specification:
\begin{equation}
p(c | y_{(1:M)}) = 
\frac{\exp\{-y_{(c)}\}}{\sum_{m=1}^M \exp\{-y_{(m)}\}}.
\label{eq:sm}
\end{equation}

The $M \times N$ latents $\mathcal{Y}$ add many degrees of freedom to
the model, enormously expanding the parameter space.  A proper prior
($a,b > 0$) for $\sigma^2_{(m)}$ is required to ensure a proper
posterior for all $N$.  There is little benefit to allowing a linear
mean function, so it is typical to take $f(x) = 0$, and thus $p=0$,
eliminating $\beta_{(m)}$ from the model.  Conditional on the
$Y_{(m)}$, samples from the posterior of the parameters to the
$m^{\mathrm{th}}$ GP may be obtained as described above via
Eq.~(\ref{eq:gpk}).  Given parameters, several schemes may be used to
sample $\mathcal{Y}$ via Eqs.~{(\ref{eq:predgp}--\ref{eq:preds2})}
[see Section \ref{sec:plgpc}].  The predictive distribution, required
for our SMC/PL algorithm, is more involved [also deferred to
Section \ref{sec:plgpc}].  Almost irrespective of the details of
implementation, inference for GP classification is much harder than
regression.  In practice, only $N \times (M-1)$ latents, and thus
$M-1$ GPs, are necessary since we may fix $Y_{(M)} = 0$, say, without
loss of generality.  Although having fewer latents makes inference a
little easier, it introduces an arbitrary asymmetry in the prior which
may be undesirable.  To simplify notation we shall use $M$ throughout,
although in our implementations we use $M-1$.

\subsection{Sequential Monte Carlo}
\label{sec:smc}

Sequential Monte Carlo (SMC) is an alternative to MCMC that is
designed for online inference in dynamic models.
In SMC, {\em particles} $\{S_t^{(i)}\}_{i=1}^N$ containing the {\em
  sufficient information} about all uncertainties given data $z^t=
(z_1,\dots,z_t)$ up to time $t$ are used to approximate the posterior
distribution: $\{S_t^{(i)}\}_{i=1}^N \sim p(S_t | z^t)$.  In Section
\ref{sec:plgp} we describe the sufficient information $S_t$ for our GP
regression and classification models.  The key task in SMC inference
is to {\em update} the particle approximation from time $t$ to time
$t+1$.

Our preferred SMC updating method is {\em particle learning} 
\citep[PL, e.g.,][]{carvalho:etal:2008} due to the convenient form of
the posterior predictive distribution of GP models.  The PL update is
derived from the following decomposition.
\begin{align*}
p(S_{t+1}|z^{t+1}) &= \int p(S_{t+1}|S_t, z_{t+1}) \;d\mathbb{P}(S_t|z^{t+1}) 
\propto  \int p(S_{t+1}|S_t, z_{t+1}) p(z_{t+1} | S_t) \;d\mathbb{P}(S_t|z^t)
\end{align*}
This suggests a two-step update of the particle approximation:
\begin{enumerate}
\item {\em resample} the indices $\{i\}_{i=1}^N$ with replacement from
  a multinomial distribution where each index has weight $w_i \propto
  p(z_{t+1} | S_t^{(i)}) = \int p(z_{t+1} | S_{t+1}) p(S_{t+1} | S_t)
  \,dS_{t+1}$, thus obtaining new indices $\{\zeta(i)\}_{i=1}^N $
\item {\em propagate} with a draw from $S_{t+1}^{(i)} \sim
  p(S_{t+1}|S_t^{\zeta(i)}, z_{t+1})$ to obtain a new collection of
  particles $\{S_{t+1}^{(i)}\}_{i=1}^N \sim p(S_{t+1}|z^{t+1})$
\end{enumerate}

The core components of PL are not new to the SMC arsenal.  Early
examples of related propagation methods include those of
\cite{kong:liu:wong:1994}, with resampling and the propagation of
sufficient statistics by \citet{liu:chen:1995,liu:chen:1998}, and
look-ahead by \citet{pitt:shep:1999}.  Like many SMC algorithms, PL is
susceptible to an accumulation of Monte Carlo error with large data
sets.  However, two aspects of our setup mitigate these concerns to a
large extent.  Firstly, the over-arching goal of sequential design is
to keep data sets as small as possible.  GPs scale poorly to large
data sets anyways, regardless of the method of inference (SMC, MCMC,
etc.), so drastically different approaches are recommended for
large-scale sequential design.  Secondly, we only use vague priors for
parameters which can be analytically integrated out in the posterior
predictive---the main workhorse of PL---so that there is no need to
sample them.  In this way we extend the class of models for which SMC
algorithms apply.  However, we note that in order to use vague priors
we must initialize the particles at some time $t_0 > 0$.  Further
explanation and development is provided in Section \ref{sec:plgp}.

\section{Particle Learning for Gaussian processes}
\label{sec:plgp}

To implement PL for GPs we need to: identify the sufficient
information $S_t$; initialize the particles; derive $p(z_{t+1}|S_t)$
for the resample step; and determine $p(S_{t+1} | S_t, z_{t+1})$ for
the propagate step.  We first develop these quantities for GP
regression and then extend them to classification.  Although GPs are
not dynamic models, we will continue to index the data size, which was
$N$ in $D_N$ in Section \ref{sec:gp}, with $t$ in the SMC framework so
that $z^t \equiv D_N$.  We use $N$ for the number of particles.  As
GPs are nonparametric priors, their sufficient information has size in
$\Omega(t)$, i.e., they depend upon the full $z^t$.  For example, the
covariance $\Sigma(X_t)$ typically requires maintaining $O(t^2)$
quantities to store the distances between the pairs of rows in $X_t$.
Therefore $z^t$ is tacitly part of the sufficient information $S_t$.

\subsection{Regression}
\label{sec:plgpr}

{\bf Sufficient Information:} Recall that $z_t = (X_t, Y_t)$ in the
regression setup.  From our discussion in Section \ref{sec:gp}, the
sufficient information, $S_t$, needed for GP regression comprises only
of the parameters of $K(\cdot, \cdot)$, defining $K_t$ via the pairs
of rows in $X_t$.  All of the other necessary quantities
($\tilde{\beta}_t \equiv \tilde{\beta}(K_t)$, and $\psi_t \equiv
\psi(K_t)$) may be calculated directly from $K_t$ and $z^t$.  However,
we prefer to think of the sufficient information as $S_t = \{K_t,
\tilde{\beta}_t, \psi_t\}$ for a clearer presentation and efficient
implementation.

\bigskip \noindent {\bf Initialization: } Particle initialization
depends upon the choice of prior for $\sigma^2$.  With a proper prior
$(a,b>0)$ we may initialize the $N$ particles at time $t_0=0$ with a
sample of the $K(\cdot, \cdot)$ parameterization from its prior,
$K^{(i)}_{0} \iidsim \pi(K)$. We then calculate
$\tilde{\beta}_{0}^{(i)},\psi_{0}^{(i)}$ from $K_{0}^{(i)}$ following
Eq.~(\ref{eq:bmean}), thereby obtaining $S_0^{(i)}$.  To take an
improper prior $(a,b=0)$ on $\sigma^2$ requires initializing the
particles conditional upon $t_0>p+1$ data points to ensure a proper
posterior.  In other words, we must start the SMC algorithm at a time
later than zero.  We find that a MH scheme for obtaining
$K_{t_0}^{(i)} \sim p(K|z^{t_0})$, via proposals from the prior
$\pi(K)$ and accepting via Eq.~(\ref{eq:gpk}), works well for $t_0$
small (i.e., not much larger than $p+1$) since the prior is similar to
the posterior ($p(K|z^{t_0})$) in this case.  We may then calculate
$\tilde{\beta}_{t_0}^{(i)},\psi_{t_0}^{(i)}$ from $K_{t_0}^{(i)}$ and
$z^{t_0}$ following Eq.~(\ref{eq:bmean}), thereby obtaining
$S_{t_0}^{(i)}$.  Both approaches (proper or improper prior for
$\sigma^2$) require a proper prior on the parameters to $K(\cdot,
\cdot)$. Sensible defaults exist for many of the typical choices for
$K(\cdot, \cdot)$.  One such choice is suggested in our illustration,
to follow shortly.

\bigskip \noindent {\bf Resample: } Technically, calculating the
weights for the resample step requires integrating over
$p(S_{t+1}|S_t)$.  But since the GP is not a dynamic model we can only
talk about $S_{t+1}$ conditional on $z_{t+1} = (x_{t+1}, y_{t+1})$.
Therefore, $p(z_{t+1} | S_t^{(i)})$ is just the probability of
$y_{t+1}$ under the Student-$t$ (\ref{eq:predgp}--\ref{eq:preds2})
given $S_t^{(i)}$: $w_t^{(i)} \propto p(y(x_{t+1}) | z^t, K_t^{(i)})
\equiv p(y(x_{t+1}) | z^t, K_t^{(i)}, \tilde{\beta}_t^{(i)},
\psi_t^{(i)})$.

\bigskip \noindent {\bf Propagate:} The propagate step updates each
resampled sufficient information $S_t^{\zeta(i)}$ to account for
$z_{t+1}=(x_{t+1}, y_{t+1})$.  Since the parameters to $K(\cdot,
\cdot)$ are static, i.e., they do not change in $t$, they may by
propagated deterministically by copying them from $S_t^{\zeta(i)}$ to
$S_{t+1}^{(i)}$.  We note that, as a matter of efficient bookkeeping,
it is the correlation matrix $K_{t+1}$ and its inverse $K_{t+1}^{-1}$
that are required for our PL update, not the values of the parameters
directly.  The new $K_{t+1}^{(i)}$ is built from $K_t^{(i)}$ and
$K^{(i)}(x_{t+1}, x_j)$, for $j=1,\dots, t+1$ as
\begin{align*}
K_{t+1}^{(i)} &= \begin{bmatrix}
K_t^{(i)} & k_t^{(i)}(x_{t+1}) \\
k_t^{(i)\top}(x_{t+1}) & K^{(i)}(x_{t+1},x_{t+1})
\end{bmatrix}.  \intertext{The partition inverse equations yield
  $(K_{t+1}^{(i)})^{-1}$ in $O(t^2)$ rather than $O(t^3)$:}
(K_{t+1}^{(i)})^{-1} &= \begin{bmatrix} [(K_t^{(i)})^{-1} +
  g_t^{(i)}(x_{t+1}) g_t^{(i)\top}(x_{t+1})/\mu_t^{(i)}(x_{t+1})]
  & g_t^{(i)}(x_{t+1}) \\
  g_t^{(i)\top}(x_{t+1}) & \mu_t^{(i)}(x_{t+1})
\end{bmatrix},
\end{align*}
where $g(x) = -\mu(x) K^{-1} k(x)$ and $\mu(x) = [K(x,x) - k^\top(x)
K^{-1} k(x)]^{-1}$.  Using Eq.~(\ref{eq:bmean}) to calculate
$\tilde{\beta}_{t+1}^{(i)}(K_{t+1}^{(i)})$ and
$\psi_{t+1}^{(i)}(K_{t+1}^{(i)})$ takes time in $O(t^2)$.  It is
possible to update these quantities in $O(\max\{t,p\})$ time
\cite[e.g.,][]{escobar:moser:1993} from their counterparts in
$S_t^{\zeta (i)}$ and the new $z_{t+1}$.  However, this would not
improve upon the overall complexity of the propagate step so we prefer
the simpler expressions (\ref{eq:bmean}).

Deterministically copying $K(\cdot, \cdot)$ in the propagate step is
fast, but it may lead to particle depletion in future resample steps.
An alternative is to augment the propagate with a sample from the
posterior distribution via MCMC to {\em rejuvenate} the particles
\citep[e.g.,][]{maceach:clyde:liu:1999,gilks:berz:2001}.  In our
regression GP context, just a single MH step for the parameters to
$K(\cdot, \cdot)$ using Eq.~(\ref{eq:gpk}), for each particle,
suffices.  The particles represent ``chains'' in equilibrium so it is
sensible to tune the MH proposals for likely acceptance by making
their variance small, initially, relative to the posterior at the
starting time $t= t_0$, and then further decreasing it
multiplicatively as $t$ increments.  Such MH rejuvenations position
the propagate step as a local maneuver in the Monte Carlo method,
whereas resampling via the predictive is a more global step.  Together
they can emulate an ensemble method.



\bigskip \noindent {\bf An illustration:} In our illustrations we
follow \cite{gra:lee:2008} and take $K(\cdot,\cdot)$ to have the form
$ K(x, x'|g) = K^*(x, x') + g \delta_{x,x'}, $ where
$\delta_{\cdot,\cdot}$ is the Kronecker delta function, and $-1 \leq
K^*(x,x') \leq 1$.  The $g$ term, referred to as the {\em nugget},
must be positive and provides a mechanism for introducing measurement
error into the stochastic process.  It causes the predictive equations
(\ref{eq:predgp}--\ref{eq:preds2}) to smooth rather than interpolate,
encoding \citep[][Appendix B]{gramacy:2005} the process $Y(x) = \mu(x)
+ \varepsilon(x) + \eta$, where $\mu$ is the mean, $\varepsilon$ is
the GP covariance structure ($\sigma^2 K^*(\cdot, \cdot)$), and $\eta$
is the noise process ($\sigma^2 g$).  We take $K^*(\cdot, \cdot)$ to
be an isotropic Gaussian correlation function with unknown {\em range}
parameter $d$: $K^*(x, x'|d) = \exp\left\{||x - x'||^{2}/d\right\}$.
Upon scaling the inputs ($X$) to lie in $[0,1]^p$ and the outputs
($Y$) to have a mean of zero and a range of one, it is easy to design
priors for $d$ and $g$ since the range of plausible values is greatly
restricted.  We use Exp$(\lambda=5)$ for both parameters throughout.
Random walk MH proposals from a uniform ``positive sliding window''
centered around the previous setting works well for both parameters.
E.g., $d^* \sim \mathrm{Unif}(\ell d/u, ud/\ell)$ for $u > \ell > 0$.
The setting $(u,\ell)=(4,3)$ is a good baseline \citep{Gramacy:2007}.
When using MH for rejuvenation one may increase $u$ and $\ell$ with
$t$ to narrow the locality.

Consider the 1-d synthetic
sinusoidal data first used by \citet{higd:2002},
\begin{equation}
    y(x) =  \sin\left(\frac{\pi x}{5}\right)
        + \frac{1}{5}\cos\left(\frac{4\pi x}{5}\right),
  \label{e:sindata}
\end{equation}
where $x\in[0,9.6]$, capturing two periods of low fidelity oscillation
(the sine term).  We observe the response with noise $Y(x) \sim
N(y(x),\sigma=0.1)$.  At this noise level it is difficult to
distinguish the high fidelity oscillations (the cosine term) from the
noise without many samples.  We used a $T=50$ Latin hypercube design
\citep[LHD, e.g.,][Section 5.2.2]{sant:will:notz:2003}---just large
enough to begin to detect the high fidelity structure.

For PL we used $N=1000$ particles with an improper, scale-invariant
prior ($a,b = 0$) for $\sigma^2$.  The particles were initialized at
time $t_0=5$ via 10,000 MH rounds, saving every $10^{\mathrm{th}}$.
This took about 30 seconds in our {\sf R} implementation on a 3GHz
Athalon workstation.  The remaining 45 PL updates with MH rejuvenation
steps ($O(t^3)$) following deterministic propagates took about 5
minutes: the first few ($t < 10$) took seconds, whereas the last few
($t > 45$) took tens of seconds.  It is this fast between--round
updating that we exploit for sequential design in Section
\ref{sec:sd}.  Foregoing rejuvenation ($O(t^2)$) drastically reduces
the computational demands for fixed $N$, but larger $N$ is needed to
get a good fit due to particle depletion.

\begin{figure}[ht!]
\centering
\includegraphics[scale=0.65,trim=10 20 10 0]{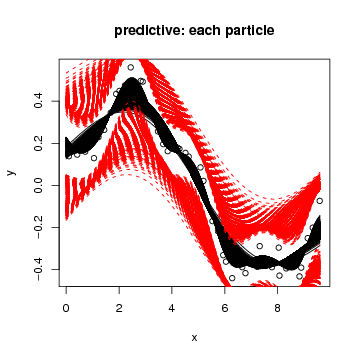} \hfill
\includegraphics[scale=0.65,trim=10 20 10 0]{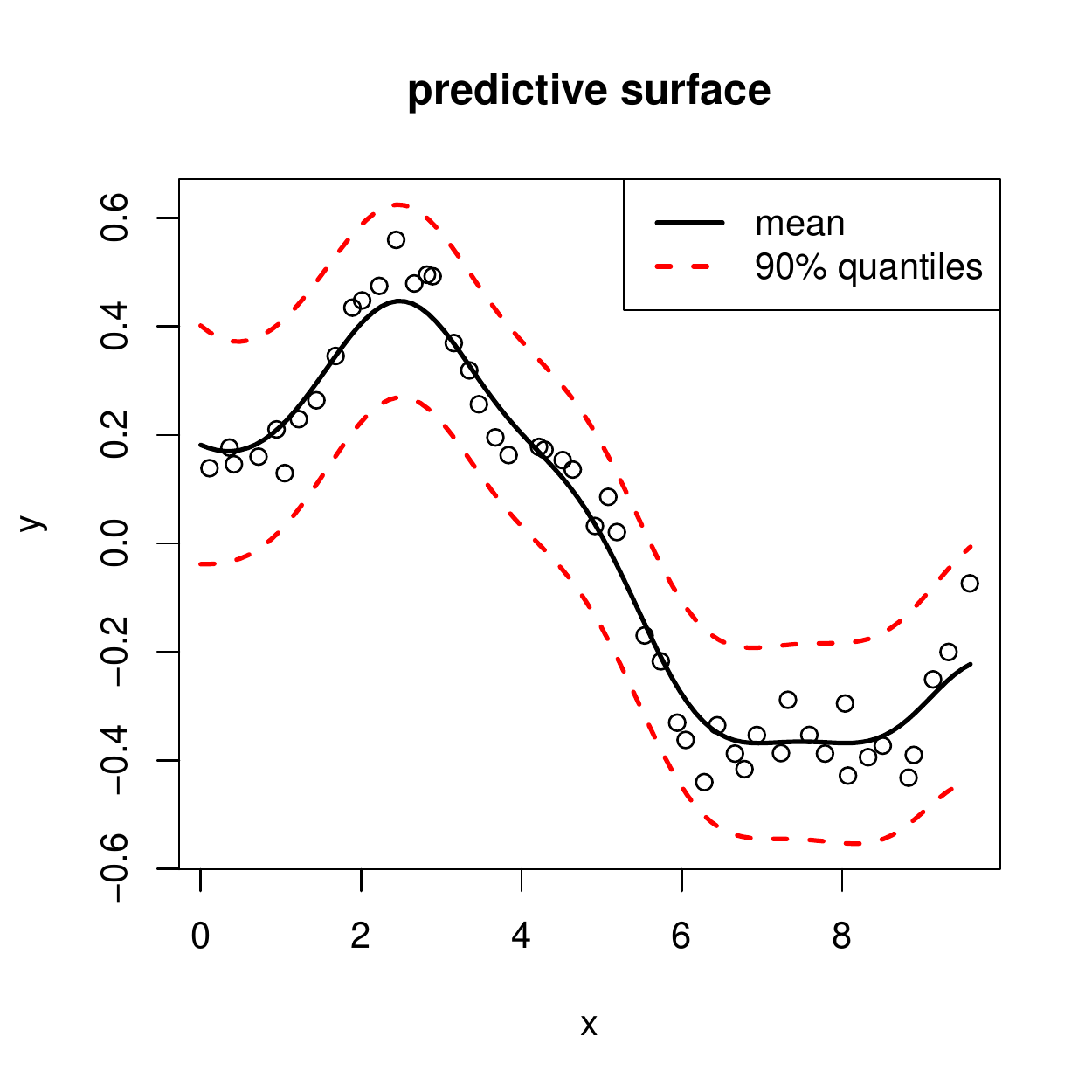}
\caption{Predictive surface(s) for the sinusoidal data in terms of the
  posterior mean (black solid) and central 90\% credible interval(s)
  (red dashed).  Each particle is is represented on the {\em left}
  with three lines, and the average of the particles is on the {\em
    right}.}
\label{f:sin}
\end{figure}

\begin{figure}[ht!]
\centering
\includegraphics[scale=0.65,trim=10 20 10 0]{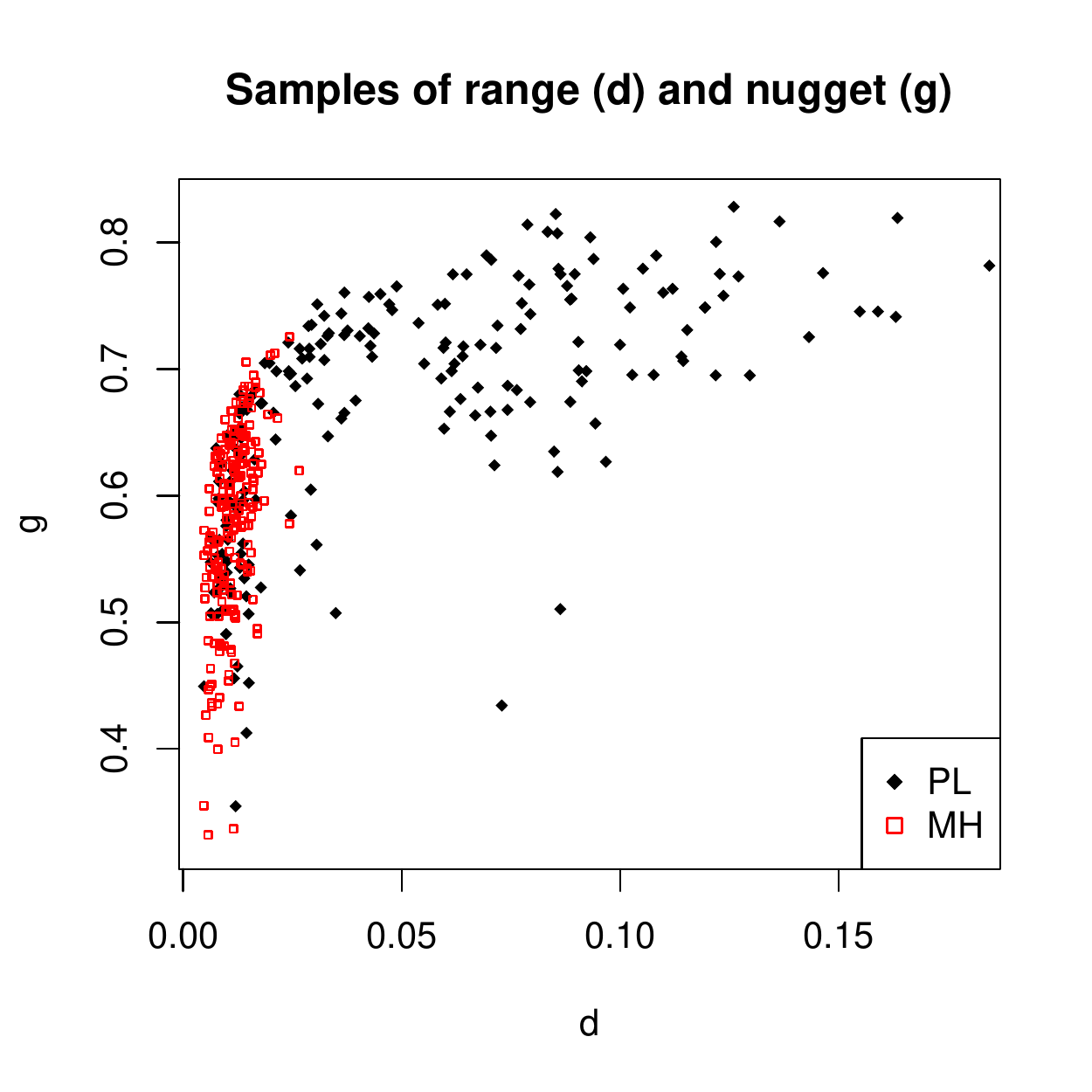}
\caption{200 samples of the range ($d$) and nugget ($g$) parameter
obtained from particles (black diamonds) and from MCMC (red squares).}
\label{f:sinsamp}
\end{figure}

The {\em left} panel of Figure \ref{f:sin} shows the point-wise
predictive distribution for each of the 1,000 particles in terms of
the mean(s) and central 90\% credible interval(s) of the Student-$t$
distributions (\ref{eq:predgp}--\ref{eq:preds2}) with parameters
$\hat{y}_t^{(i)}$, $\hat{\sigma}^{2(i)}_t$ and $\hat{\nu}_t^{(i)}$
obtained from $S_t^{(i)}$.  Their average, the posterior mean
predictive surface, is shown on the {\em right}.  Observe that some
particles lead to higher fidelity surfaces (finding the cosine) than
others (only finding the sine).  Figure \ref{f:sinsamp} shows the
samples of the range ($d$) and nugget ($g$) obtained from the
particles.  Only 200 of the 1,000 are shown to reduce clutter.  The
clustering pattern of the black diamonds indicates a multimodal
posterior.

For contrast we also took 10,000 MCMC samples from the full data
posterior, thinning every 10 and saving 1,000.  This took about one
minute on our workstation, which is faster than the full PL run, but
much slower than the individual updates $t\rightarrow t+1$.  The
marginal chains for $d$ and $g$ seemed to mix well (not shown) but, as
Figure \ref{f:sinsamp} shows [plotting last 200 sample pairs as red
squares], the chain nevertheless became stuck in a mode of the
posterior, and only explored a portion of the high density region.

For a more numerical comparison we calculated the RMSE of predictive
means (obtained via PL and MCMC, as above) to the truth on a random
LHD of size 1000.  This was repeated 100 times, each with new LHD
training (size 50, as above) and test sets.  The mean (sd) RMSE was
$0.00079 \; (0.00069)$ for PL, and $0.00098 \; (0.00075)$ for MCMC.
As paired data, the average number of times PL had a lower RMSE than
MCMC was 0.64, which is statistically significant ($p = 5.837\times
10^{-5}$) using a standard one-sided $t$-test.  In short, this means
that the SMC/PL method is performing at least as well as the MCMC with
quicker sequential updates.  The MCMC could be re-tuned, restarted,
and/or run for longer to narrow the RMSE gap, but all of these would
come at greater computational expense.

\subsection{Classification}
\label{sec:plgpc}

{\bf Sufficient Information: } In classification we use $M$ GP priors
on $M \times t$ latent variables.  Therefore, $S_t$ comprises of
$\{K_{(m),t}, \tilde{\beta}_{(m),t}, \psi_{(m),t}\}_{m=1}^M$ and
$\mathcal{Y}^t$.  

\bigskip \noindent {\bf Initialization: } Particle initialization is
identical to an $M$--fold application of regression GP particle
initialization.  As remarked in Section \ref{sec:gp}, we must use a
proper prior $(a,b > 0)$ for each $\sigma_{(m)}^2$.  As a consequence,
we may initialize all of the particles at $t_0 = 0$ by sampling
$\{K_{(m),0}^{(i)}\}_{m=1}^M$ identically from $\pi(K)$.  There are no
latent $\mathcal{Y}$ at time zero, so neither they nor
$\{\tilde{\beta}_{(m),0}, \psi_{(m),0}\}_{m=1}^M$ are required.  It is
also possible to initialize the particles at $t_0 > 0$, which may be
desirable in some situations.  In this case, a hybrid of the MH scheme
for regression GP's, applied $M$--fold, and a sampling of the latent
$\mathcal{Y}^{t_0}$ yielding $\{\tilde{\beta}_{(m),t_0},
\psi_{(m),t_0}\}_{m=1}^M$, as described below for the propagate step,
works well.

\bigskip \noindent {\bf Resample:} It may be helpful to think of the
latent $\mathcal{Y}^t$ as playing the role of (hidden) states in a
dynamic model.  Indeed, their treatment in the PL update is similar.
However, note that they do not satisfy any Markov property.  The
predictive density $p(z_{t+1} | S_t)$, which is needed for the
resample step, is the probability of the label $c_{t+1}(x_{t+1})$
under the sufficient information $S_t$: $p(c_{t+1}(x_{t+1}) | S_t)$.
This depends upon the $M$ latents $\mathcal{Y}(x_{t+1})$, which are
not part of $S_t$.  For an arbitrary $x$, the law of total probability
gives
\begin{align}
p(c(x) | S_t)
&= \int_{\mathbb{R}^M} p(c(x), \mathcal{Y}(x)| S_t) \;
d \mathcal{Y}(x) \label{eq:cpred} 
= \int_{\mathbb{R}^M} p(c(x) | \mathcal{Y}(x))
p(\mathcal{Y}(x) | S_t) \; d \mathcal{Y}(x). 
\end{align}
The second equality comes since,
 conditional on $\mathcal{Y}(x)$, the label 
does not depend on any other quantity
(\ref{eq:sm}).  The $M$ GP priors are independent, 
so $p(\mathcal{Y}(x) | S_t)$ decomposes as
\begin{equation}
p(\mathcal{Y}(x) | S_t) = \prod_{m=1}^M p(y_{(m)}(x) |
Y_{(m),t}, K_{(m),t}),
\label{eq:tprod}
\end{equation}
where each component in the product is a Student-$t$ density
(\ref{eq:predgp}--\ref{eq:preds2}).

The $M$-dimensional integral in Eq.~(\ref{eq:cpred}) is not
analytically tractable, but it is trivial to approximate
by Monte Carlo as follows.  Simulate many independent collections of 
samples from each of the $M$ Student-$t$ distributions (\ref{eq:tprod}):  
\begin{align}
  \tilde{Y}(x)^{(\ell)} &= \{\tilde{y}_{(m)}(x)^{\ell}\}_{i=1}^M, &&
  \mbox{where}  & \tilde{y}_{(m)}(x)^{\ell} &\iidsim 
  p(y_{(m)}(x) | Y_{(m),t}, K_{(m),t}),
\label{eq:ypredl}
\end{align}
for $\ell = 1,\dots, L$, say---thereby collecting $M\times L$ samples.
Then pass these latents through the likelihood (\ref{eq:sm}) and
take an average:
\begin{equation}
p(c(x) | S_t) \approx 
\frac{1}{L} \sum_{\ell = 1}^L p(c(x) | \tilde{Y}(x)^{(\ell)}).
\label{eq:cpredl}
\end{equation}
With as few as $L=100$ samples this approximation is quite accurate.
The weights $\{w_i\}_{i=1}^N$, where $w_i \propto p(c_{t+1}(x_{t+1}) |
S_t^{(i)})$, may be used to obtain the resampled indices
$\{\zeta(i)\}_{i=1}^N$.  Observe that even $L=1$ is possible, since we
may then view $\tilde{Y}(x)^{(1)}$ as an auxiliary member of the state
$S_t$.  So any inaccuracies in the approximation simply contribute to
the Monte Carlo error of the PL method, which may be squashed with
larger $N$.

\bigskip \noindent {\bf Propagate:} The GP classification propagate
step is essentially the aggregate of $M$ regression propagates.  But
these may only commence once the new latent(s) for $t+1$ are
incorporated.  We may extend the hidden state analogy to sample
$\mathcal{Y}_{:,t+1}^{\zeta(i)} \sim p(\mathcal{Y}(x_{t+1}) |
S_t^{\zeta(i)})$ via the independent Student-$t$ distributions
(\ref{eq:tprod}).  In expectation we have that
$\{\tilde{\beta}_{(m),t}^{\zeta(i)},\psi_{(m),t}^{\zeta(i)}\}_{m=1}^M
=
\{\tilde{\beta}_{(m),t}^{\zeta(i)},\psi_{(m),t}^{\zeta(i)}\}_{m=1}^M|\mathcal{Y}^{t+1,\zeta(i)},
\{K_{(m),t}^{\zeta(i)}\}_{m=1}^M$ since $\mathcal{Y}_{t+1}^{\zeta(i)}
\sim p(\mathcal{Y}(x_{t+1}) | S_t^{\zeta(i)})$, so no update of the
rest of the sufficient information is necessary at this juncture.  To
complete the propagation we must sample the full set of latents
$\mathcal{Y}^{t+1,\zeta(i)}$ conditional upon $c_{t+1}$ via $z^{t+1}$.
Once obtained, these fully propagated latents $\mathcal{Y}^{t+1,(i)}$
may be used to update the remaining components of the sufficient
information: $\{\tilde{\beta}_{(m),t+1}^{(i)},\psi_{(m),t+1}^{(i)},
K_{(m),t+1}^{(i)}\}_{m=1}^M| \mathcal{Y}^{t+1,(i)}$ comprising
$S_{t+1}^{(i)}$.

Sampling the latents may proceed via ARS, following \cite{neal:1998}.
However, as in the regression setup, we prefer a more local move in
the PL propagate context to compliment the globally-scoped resample
step.  So instead we follow \cite{brod:gra:2010} in using 10-fold
randomly blocked MH-within-Gibbs sampling.  This approach exploits a
factorization of the posterior as the product of the class likelihood
(\ref{eq:sm}) given the underlying latents and their GP prior
(\ref{eq:cpred}): (dropping the $\zeta(i)$)
\begin{equation}
p(C(X_I) | \mathcal{Y}^{t+1}(X_I)) \times 
p(Y_{(m)}^{t+1}(X_I)| Y^{t+1}_{(m)}(X_{-I}), K(\cdot, \cdot)). \label{eq:decomp}
\end{equation}
Here, $I$ is an element of a 10-fold (random) partition
$\mathcal{I}_{10}$ of the indices $1,\dots, t+1$, where $|I| \leq 10$
and $-I = \mathcal{I}_{10} \backslash I$ is its compliment.  Extending
the predictive equations from Section \ref{sec:plgpr}, the latter term
in Eq.~(\ref{eq:decomp}) is an $|I|$-dimensional Student-$t$ with
$\hat{\nu}_I = |\!-\!I|-p-1$,
\begin{align}
\mbox{mean vector} && \hat{Y}_I &= F_I \tilde{\beta}_{-I} + 
K_{I,-I}, K_{-I,-I}^{-1} (Y_{-I} - F_{-I}, \tilde{\beta}_{-I}), \\
\mbox{and scale matrix} && \hat{\Sigma}_{I,I} &=
\frac{(b + \psi_{-I})[K_{I,-I} - K_{I,-I} K_{I,I}^{-1} K_{-I,I}]}{
a + \hat{\nu}_I}, \nonumber
\end{align}
using the condensed notation $Y_I \equiv Y_{(m)}(X_I)$, and $|I|
\times |I'|$ matrix $K_{I,I'} \equiv K_{(m)}(X_I, X_{I'})$, etc. A
thus proposed $Y_{(m)}'(X_I)$ may be accepted according to the
likelihood ratio since the prior and proposal densities cancel in the
MH acceptance ratio.  Let $\mathcal{Y}'_I$ denote the collection of $M
\times (t+1)$ latents comprised of $Y_{(m)}'(X_I)$,
$Y_{(-m)}^{t+1}(X_I)$, and $\mathcal{Y}^{t+1}(X_{-I})$.  Then the MH
acceptance probability is $\min\{1,A\}$ where
\[
A = \frac{p(C(X_I) | \mathcal{Y}'_I)}{p(C(X_I) | \mathcal{Y}_I^{t+1})}
= \prod_{i\in I} \frac{p(c_i | \mathcal{Y}'_i)}{p(c_i | \mathcal{Y}_i^{t+1})}.
\]
Upon acceptance we replace $Y_{(m)}^{t+1}(X_I)$ with $Y_{(m)}'(X_I)$,
and otherwise do nothing.  In this way we loop over $m=1,\dots,M$ and
$I \in \mathcal{I}_{10}$ to obtain a set of fully propagated latents.



\bigskip \noindent {\bf An Illustration:} Consider data generated by
converting a real-valued output $y(x) = x_1 \exp(-x_1^2 - x_2^2)$ into
classification labels \citep{brod:gra:2010} by taking the sign of the
sum of the eigenvalues of the Hessian of $y(x)$.  This gives a
two-class process where the class is determined by the direction of
concavity at $x$.  For our illustration we take $x\in[-2,2]^2$, and
create a third class from the first class (negative sign) where $x_1 >
0$.  We use $M-1=2$ GPs, and take our data set to be $T=125$
input--class pairs from a maximum entropy design \citep[MED, ][Section
6.2.1]{sant:will:notz:2003}.  Our $N=1000$ particles are initialized
using 10,000 MCMC rounds at time $t_0 =17$, thinning every 10.  This
takes less than 2 minutes in {\sf R} on our workstation.  Then we
proceed with 108 PL updates, which takes about four hours.  The first
few updates take less than a minute, whereas the last few take 7--8
minutes.

\begin{figure}[ht!]
\centering
\includegraphics[scale=0.65,trim=10 20 10 0]{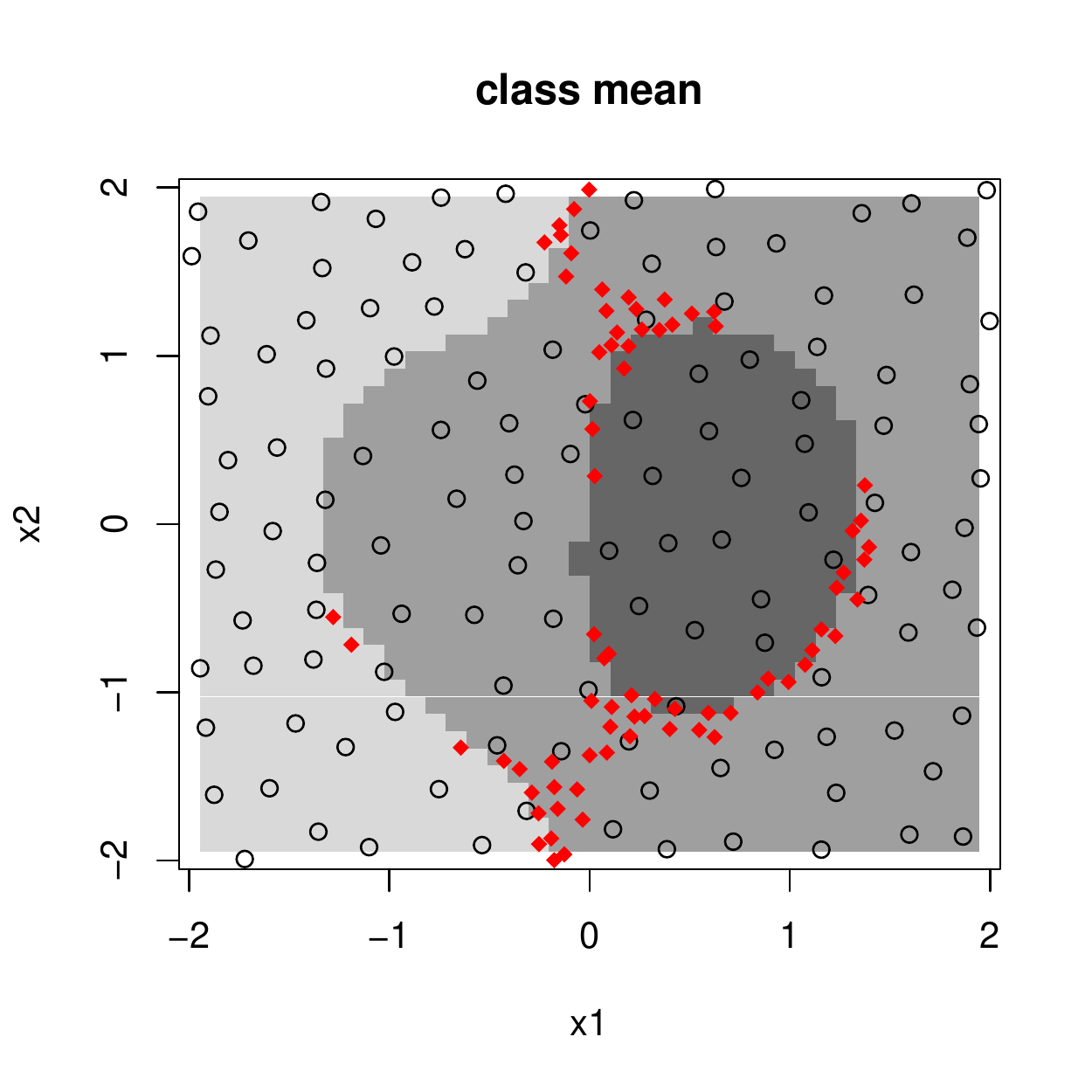} \hfill
\includegraphics[scale=0.65,trim=10 20 10 0]{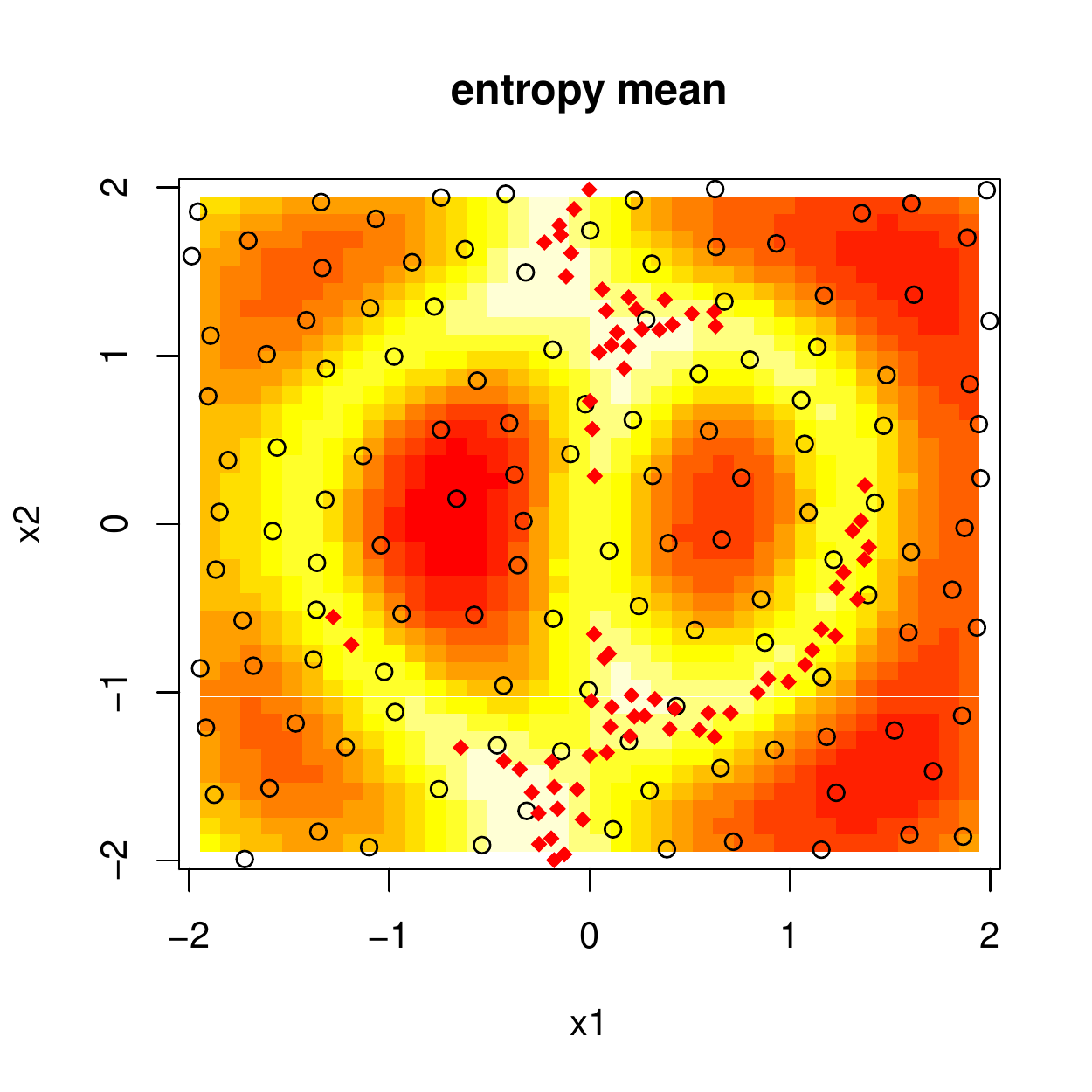}
\caption{Class posterior mean ({\em left}) and entropy posterior mean 
({\em right}) for the PL fit to the 3-class 2-d exponential data.
The classes are represented by three shades of gray; and the heat map
for the entropy is hottest (whitest) for large values.  The inputs are
black open circles, and the miss-classified predictive locations are
solid red circles.}
\label{f:class}
\end{figure}

Figure \ref{f:class} shows the posterior predictive surface,
interpolated from 1,000 MED test locations, in terms of the most
likely label from the mean posterior predictive ({\em left}), i.e.,
$\mathrm{arg}\max_m N^{-1} \sum_{i=1}^N p_m^{(i)}(x)$ where
$p_m^{(i)}(x) \equiv p(c(x) = m)^{(i)} \approx p(C(x)=m | S_t^{(i)})$,
and the mean entropy ({\em right}) of the label distribution
$-\sum_{m=1}^M p_m^{(i)}(x) \log p_m^{(i)}(x)$.  The 125 training
inputs are shown as open black circles and the 76 misclassified test
locations are shown as solid red ones.  Observe that the predictive
entropy is highest where determining the class label is most
difficult: near the boundaries.

The differences in Monte Carlo efficiency between PL and MCMC, here,
are less stark.  There is less scope for the posterior to be
multimodal due to the role of the nugget.  For classification, the
nugget parameterizes the continuum between logit (small nugget) and
probit (large nugget) models \citep{neal:1998}, which is a far more
subtle than interpolation versus smoothing as in regression.  In terms
of computational complexity we can offer the following comparison.
Obtaining 10,000 MCMC samples, thinning every 10, for the full $T=125$
input--class pairs took about 45 minutes.  While this is several times
faster than PL on aggregate, observe that a single PL update for the
$126^{\mathrm{th}}$ input--class pair can be performed several times
faster than running a full MCMC from scratch.

For a further comparison of timings on a larger classification problem
we duplicated the 10-fold cross validation (CV) experiment of
\cite{brod:gra:2010} on the two-class credit approval data which has
$p=47$ covariates for 690 $(x,c)$ pairs.  The time required for the
final PL update ($t\approx 621$) with $N=1000$ particles, averaged
over the 10 CV folds, was 38 minutes.  The resulting predictor(s) gave
exactly the same misclassification error(s) averaging $14.6\%$ ($4\%$
sd) on the hold out sets as a similar estimator based on MCMC.
However, the authors reported that the MCMC took about $5.5$ hours on
average.  So even with a modestly large design ($\approx 621$),
the Monte Carlo error that might accumulate with the use of vague
priors in SMC does not seem to (yet) be an issue in our PL
implementation. The savings in time is huge due the decomposition of
far fewer $621 \times 621$ covariance matrices in the SMC framework.

\section{Sequential design}
\label{sec:sd}

Here, we illustrate how the online nature of PL is ideally suited to
sequential design by AL.  Probably the most straightforward AL
algorithms in the regression context are ALM and ALC [see Section
\ref{sec:gp}].  But these are well known to approximate space filling
MEDs for stationary GP models.  So instead we consider the sequential
design problem of optimizing a noisy black box function.
In the classification context we consider the sequential
exploration of classification boundaries. 

\subsection{Optimization by expected improvement}
\label{sec:opt}

\cite{jones:schonlau:welch:1998} described how to optimize a
deterministic black box function using a ``surrogate'' model (i.e., a
GP with $g=0$) via the MLE (for $\{d, \beta, \sigma^2\}$).  The
essence is as follows.  After $t$ samples are gathered, the current
minimum is $f_{\min,t} = \min\{y_1,\dots, y_t\}$.  The improvement at
$x$ is $I_t(x) = \max\{f_{\min,t} - Y_t(x), 0\}$, a random variable
whose distribution is determined via $Y_t(x) \equiv Y(x)|z^t, K_t$,
which has a Student-$t$ distribution
(\ref{eq:predgp}--\ref{eq:preds2}).  The {\em expected improvement}
(EI) is obtained by analytically integrating out $Y_t(x)$.  A branch
and bound algorithm is then used to maximize the EI to obtain the next
design point $x_{t+1} = \mathrm{arg} \max \bE\{I_t(x)\}$.  The
resulting iterative procedure (choose $x_{t+1}$; obtain
$y_{t+1}(x_{t+1})$; refit and repeat) is called the {\em efficient
  global optimization} (EGO) algorithm.

The situation is more complicated when optimizing a noisy function, or
with Bayesian inference via Monte Carlo.  A re-definition of
$f_{\min,t}$ accounts for the noisy ($g>0$) responses: either as the
first order statistic of $Y(X_t)$ or as the minimum of the predictive
mean surface, $\min_x \hat{y}_t(x)$.  Now, each sample (e.g., each
particle) from the posterior emits an EI.  Using our Student-$t$
predictive equations (\ref{eq:predgp}--\ref{eq:preds2}) for
$S_t^{(i)}$, letting $\delta_t^{(i)}(x) = f_{\min,t} -
\hat{y}^{(i)}_t(x)$, we have \citep[following][]{will:sant:notz:2000}:
\begin{equation}
\bE\{I_t(x)|S_t^{(i)}\} = \delta_t^{(i)}(x) T_{\hat{\nu}_t^{(i)}}\!
\left(\frac{\delta_t^{(i)}(x)}{\hat{\sigma}_t^{(i)}(x)} \right) +
\frac{1}{\hat{\nu}_t^{(i)} - 1}
\left[\hat{\nu}_t^{(i)}\hat{\sigma}_t^{(i)}(x) +
  \frac{\delta_t^{(i)}(x)^2}{\hat{\sigma}_t^{(i)}(x)} \right]
t_{\hat{\nu}_t^{(i)}}\!
\left(\frac{\delta_t^{(i)}(x)}{\hat{\sigma}_t^{(i)}(x)}\right).
\label{eq:ei}
\end{equation}
The EI is then approximated as $\bE\{I_t(x)\} \approx N^{-1}
\sum_{i=1}^N \bE\{I_t(x)|S_t^{(i)}\}$, thereby taking parameter
uncertainty into account.  But the branch and bound algorithm no
longer applies.

A remedy, proposed to ensure convergence in the optimization, involves
pairing EI with a deterministic numerical optimizer.
\cite{tadd:lee:gray:grif:2009} proposed using a GP/EI based approach
(with MCMC) as an oracle in a pattern search optimizer called APPS.
This high powered combination offers convergence guarantees, but
unfortunately requires a highly customized implementation that
precludes its use in our illustrations.  \citet[][Section
3]{gramacy:taddy:2009} propose a simpler, more widely applicable,
variant via the opposite embedding.  There are (as yet) no convergence
guarantees for this heuristic, but it has been shown to perform well
in many examples.

Both methods work with a fresh set of random candidate locations
$\tilde{X}_t$ at each time $t$, e.g., a LHD.  In the oracle approach,
the candidate which gives the largest EI, $x_t^* = \mathrm{arg}
\max_{\tilde{x} \in \tilde{X}_t} \bE\{I_t(\tilde{x})\}$, is used to
augment the search pattern used by the direct optimizer (APPS) to find
$x_{t+1}$.  In the simpler heuristic approach the candidate design is
augmented to include the minimum mean predictive location based upon
the MAP parameterization at time $t$.  In our SMC/PL implementation
this involves first finding $i^* = \mathrm{arg} \max_{i=1,\dots, N}
p(S_t^{(i)}|z^t)$, and then finding $x_t^* = \mathrm{arg} \min_x
\hat{y}_t^{(i^*)}(x)$.  ({\sf R}'s {\tt optim} function works well for
the latter search when initialized with $\mathrm{arg} \min_{x\in
  \tilde{X}_t} \hat{y}_t^{(i^*)}(x)$.)  We may then take $x_{t+1} =
\mathrm{arg} \max_{\tilde{x} \in \tilde{X}_t\cup x_t^*}
\bE\{I_t(\tilde{x})\}$, having searched both globally via
$\tilde{X}_t$, and locally via $x_t^*$.

\begin{figure}[ht!]
\centering
\includegraphics[scale=0.65,trim=10 20 10 0]{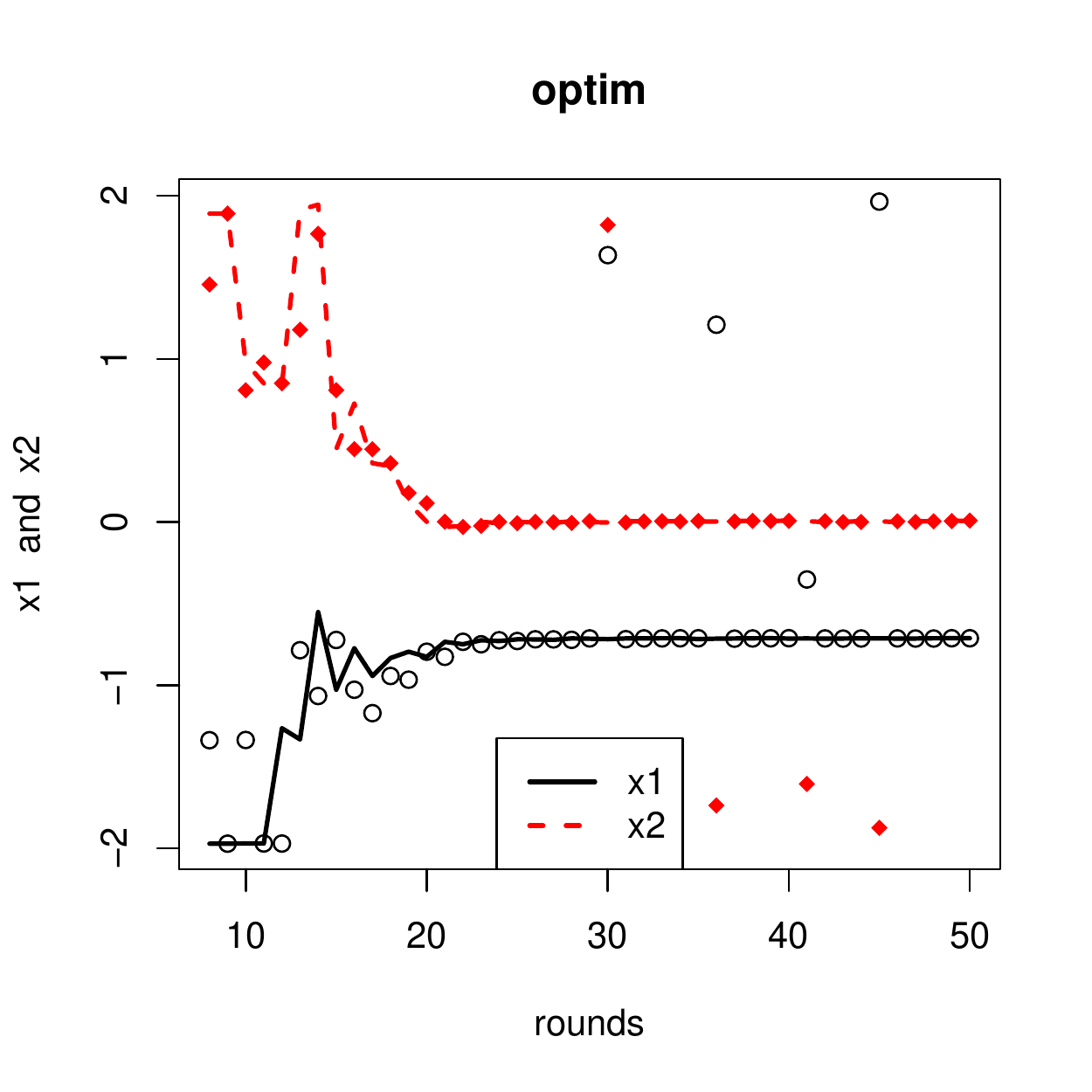} \hfill
\includegraphics[scale=0.65,trim=10 20 10 0]{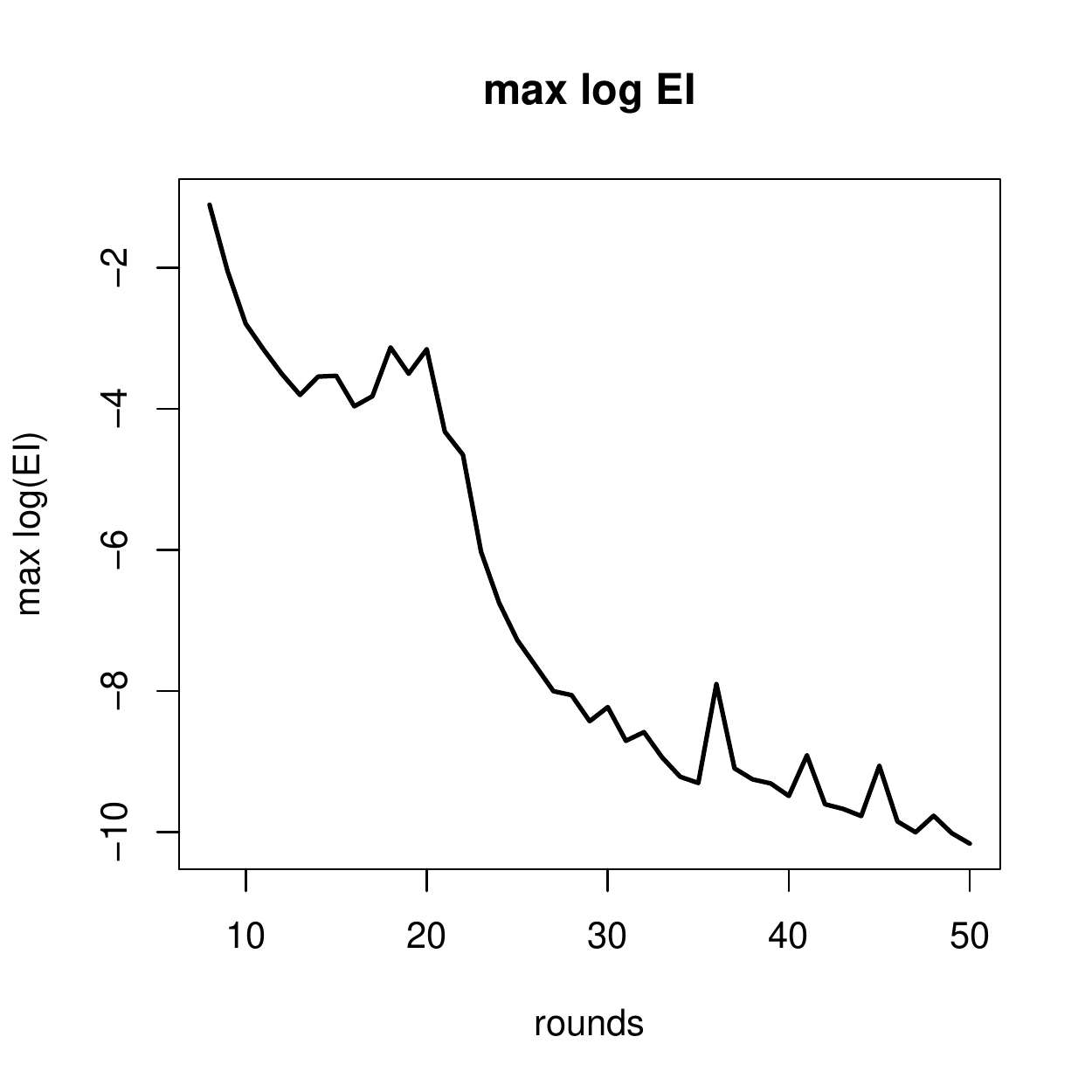}
\caption{Tracking the progress of GP/EI optimization via PL.  The
{\em left} plot shows $x_t$ (points) and $x_t^*$ (lines); the {\em right}
plot shows $\log \bE\{I_t(x_{t+1})\}$.}
\label{f:ei}
\end{figure}

Figure \ref{f:ei} illustrates the progress of this algorithm with PL
inference on the 2-d exponential data [Section \ref{sec:plgpc}],
observed with $N(0,\sigma = 0.001)$ noise.  The $N=1000$ particles
were initialized at time $t_0=7$ with a LHD.  Each $\tilde{X}_t$ is a
fresh size 40 LHD.  The {\em left} panel tracks $x_t^* = (x^*_{1,t},
x^*_{2,t})$, the optimal additional candidate, as lines and the chosen
$x_{t+1} = (x_{1,t+1}, x_{2,t+1})$ as points, both from
$t=t_0,\dots,T=50$.  Observe how the points initially explore to find
a (local) optima, and then later make excursions (unsuccessfully) in
search of an alternative.  The {\em right} panel tracks the maximum of
the log EI, $\log(\bE\{I_t(x_{t+1})\})$, from $t=t_0,\dots, T$.
Observe that this is decreasing except when $x_{t+1} \ne x_t^*$,
corresponding to an exploration event.  The magnitude and frequency of
these up-spikes decrease over time, giving a good empirical
diagnostic of convergence.  At the end we obtained $ x_T^* = (-0.7119,
0.0070)$, which is very close to the true minima $x ^*=
(-\sqrt{1/2},0)$.  The $43$ PL updates, with searches, etc., took
about eleven minutes in {\sf R} on our workstation.  By way of
comparison, the equivalent MCMC-based implementation (giving nearly
identical results) took more than 45 minutes.

\subsection{Online learning of classification boundaries}

In Section \ref{sec:plgpc} [Figure \ref{f:class}] we saw how the
predictive entropy could be useful as an AL heuristic for boundary
exploration.  \cite{joshi:2009} observed that when $M > 2$, the
probability of the irrelevant class(es) near the boundary between two
classes can influence the entropy, and thus the sequential design
based upon it, in undesirable ways.  They showed that restricting the
entropy calculation to the two highest probabilities ({\em
  best--versus--second--best} [BVSB] entropy) is a better
heuristic. 

\begin{figure}[ht!]
\centering
\includegraphics[scale=0.65,trim=10 20 10 0]{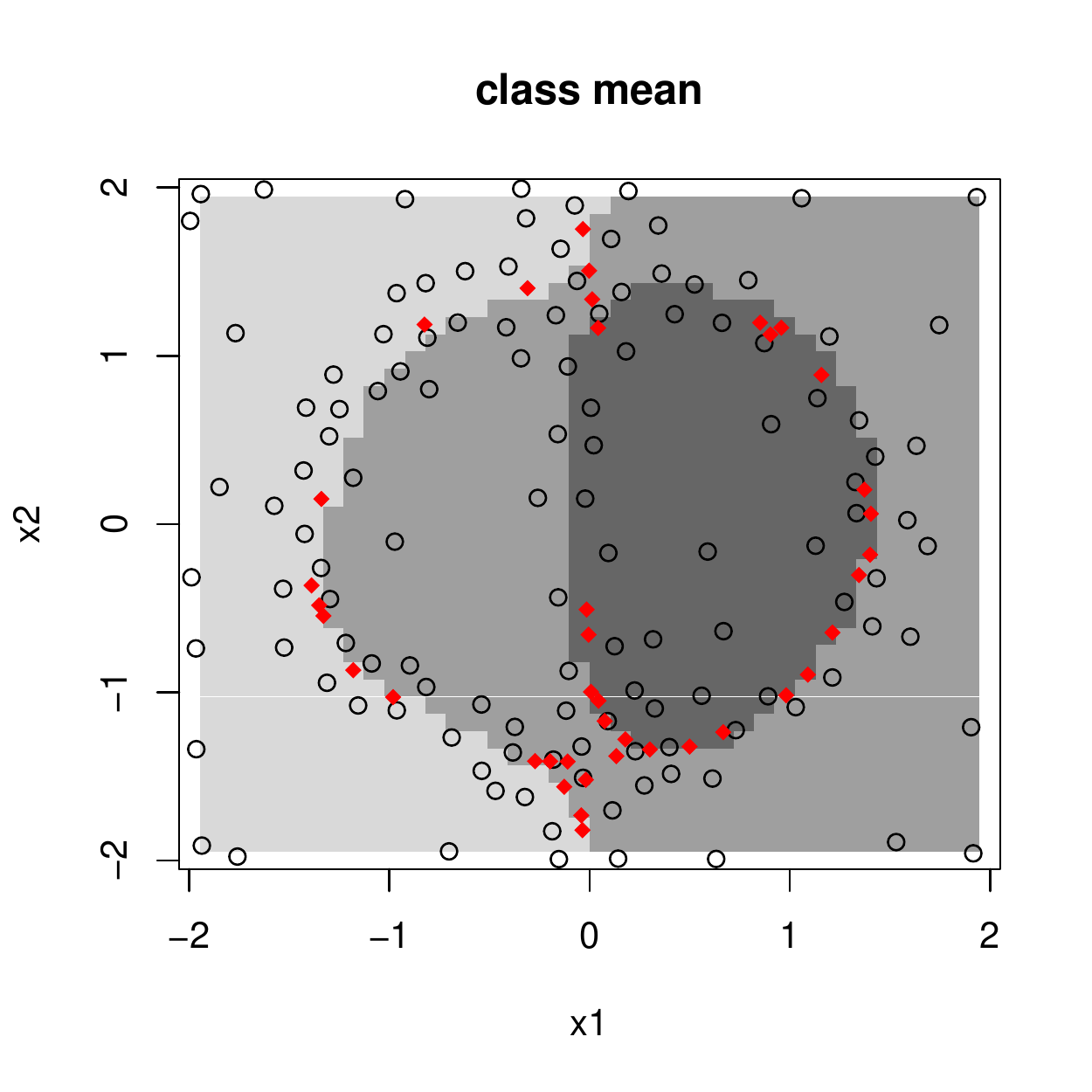} \hfill
\includegraphics[scale=0.65,trim=10 20 10 0]{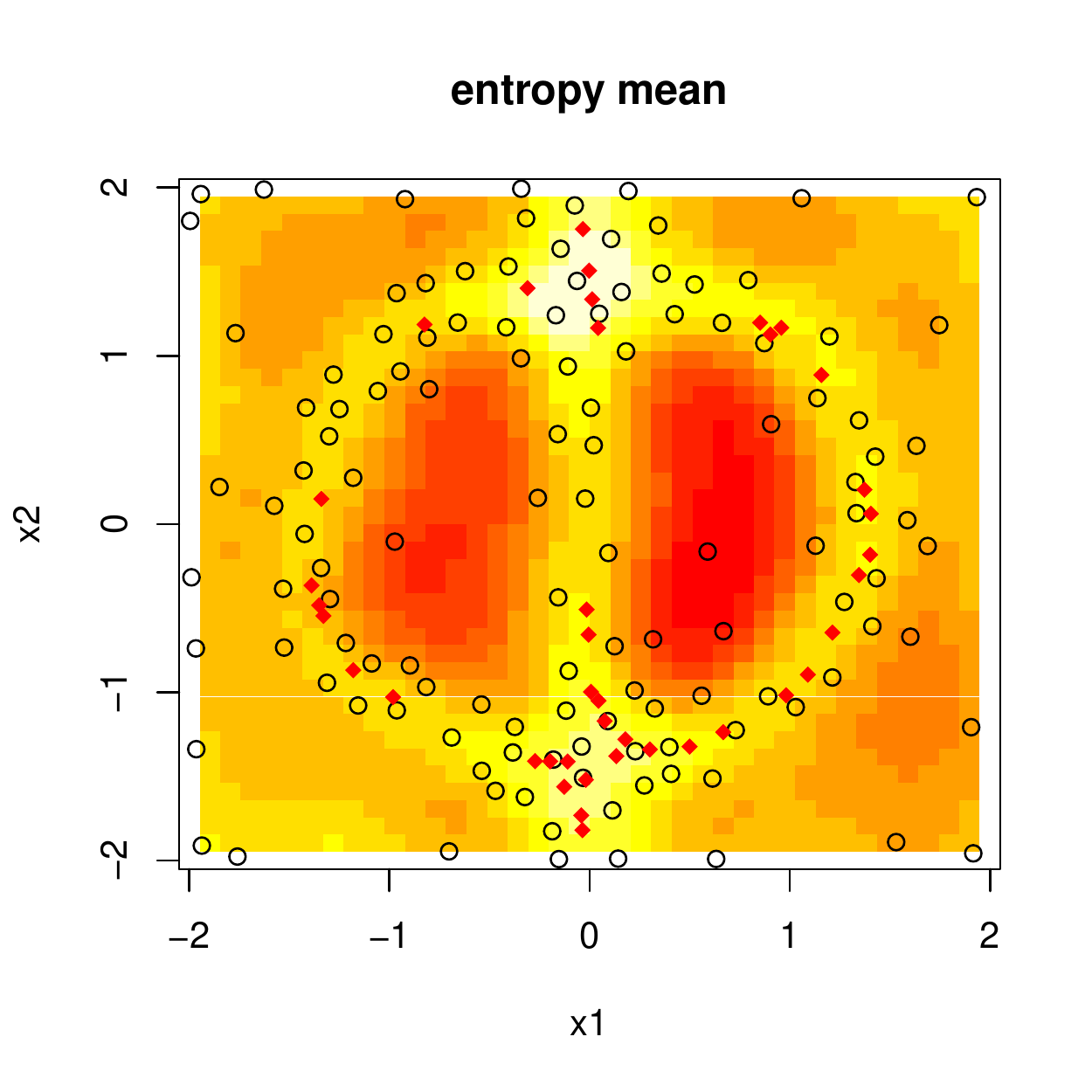}
\caption{Class posterior mean ({\em left}) and entropy posterior mean 
({\em right}) for the PL fit to the 3-class 2-d exponential data by
AL with the BVSB entropy heuristic, for comparison
with the static design version in Figure \ref{f:class}.}
\label{f:classas}
\end{figure}

Figure \ref{f:classas} shows the sequential design obtained via PL
with $N=1000$ particles and the BVSP entropy AL heuristic using a
pre-defined set of 300 MED candidate locations.  The design was
initialized with a $t_0 = 25$ sub-MED (from the 300), and AL was
performed at each of rounds $t=t_0, \dots, T=125$ on the $300-t$
remaining candidates.  This time there are 40 misclassified points,
compared to the 76 obtained with a static design [Section
\ref{sec:plgpc}; the same 1,000 MED test set was used].
The running time here is comparable to the static implementation.
MCMC gives similar results but takes 4--5 times longer.

Working off-grid, e.g., with a fresh set of LHD candidates in each AL
round, is slightly more challenging because the predictive entropy is
very greedy.  Paradoxically, the highest (BVSB) entropy regions tend
to be near the boundaries which have been most thoroughly
explored---straddling it with a high concentration of points---even
though the entropy rapidly decreases nearby.  One possible remedy
involves smoothing the entropy by a distance-based kernel (e.g.,
$K(\cdot,\cdot)$ from the GP) over the candidate locations.
Applying this heuristic leads to very similar results as those
reported in Figure~\ref{f:classas}, and so they are not shown here.

\section{Discussion}
\label{sec:discuss}

We have shown how GP models, for regression and for classification,
may be fit via the sequential Monte Carlo (SMC) method of particle
learning (PL).  We developed the relevant expressions, and provided
illustrations on data from both contexts.  Although SMC methods are
typically applied to time series data, we argued that they are also
well suited to scenarios where the data arrive online even when there
is no time or dynamic component in the model. Examples include
sequential design and optimization, where a significant aspect of the
problem is to choose the next input and subsequently update the model
fit.  In these contexts, MCMC inference has reigned supreme.  But MCMC
is clearly ill-suited to online data acquisition, as it must be
restarted when the new data arrive.  We showed that the PL update of a
particle approximation is thrifty by contrast, and that adding
rejuvenation to the propagate steps mimicks the behavior of an
ensemble without explicitly maintaining one.

Another advantage of SMC methods is that they are ``embarrassingly
parallelizable'', since many of the relevant calculations on the
particles may proceed independently of one another, up to having a
unique computing node for each particle.  In contrast, the Markov
property of MCMC requires that the inferential steps, to a large
extent, proceed in serial.  Getting the most mileage out of our SMC/PL
approach will require a careful asynchronous implementation.  Observe
that the posterior predictive distribution, and the propagate step,
may be calculated for each particle in parallel.  Resampling requires
that the particles be synchronized, but this is fast once the particle
predictive densities have been evaluated.  Our implementation in the
{\tt plgp} package does not exploit this parallelism.  However, it
does make heavy use of {\sf R}'s {\tt lapply} method, which
automatically loops over the particles to calculate the predictive,
and to propagate.  A parallelized {\tt lapply}, e.g., using {\tt
  snowfall} and {\tt sfCluster}, as described by \cite{knaus:2009},
may be a promising way forward.

\bibliography{plgp}
\bibliographystyle{jasa}

\end{document}